\documentstyle[psfig]{article}

\begin{document}
\



\noindent
{\LARGE {\bf Three steps in intranuclear cascading}}
\vspace{.5cm}

\noindent
{\Large Tadeusz Wibig}
\vspace{.3cm}

{
\large
\noindent
{ \it Experimental Physics Dept., University of \L odz, \\
Pomorska 149/153, 90-236 \L odz, Poland}
\vspace{.5cm}



\noindent
{\bf Abstract}

\noindent
{Between the initial excitation of nucleons by the passing throughout each
other and the final cascading of newly created hadrons there should be
considered one more step.
We call it the second step cascading process.
It acts as an interaction of wounded nucleon from one
nucleus with another nucleon from the same nucleus.
This mechanism leads to the significant increase on the
inelasticity for nuclei collisions without changing the hadron--hadron
interaction characteristics.}
\vspace{.3cm}

\vspace{.2cm}

The nucleus--nucleus interaction can be described quite successfully within
the framework of wounded nucleon model. It can
be expressed in the statement that once one of projectile nucleons
interacts inelastically with the one from a target nucleus
intermediate states called ``wounded nucleons'' are
created. The spatial extension of the wounded nucleon is the same as
original nucleon before the collision. The subsequent
collisions inside the target nucleus take place before this ``excited state''
hadronize.

The detailed insight into the intranuclear collisions process gave
a surprising result which lead to the revision of, e.g., the conventional
interpretation of connections between inelasticity in nuclei and hadronic
collisions and respective cross section ratios \cite{gmcinel}.
In the wounded nucleons picture of the high-energy nucleus--nucleus
interaction the non-zero time interval between excitation and hadronization
leads to the possibility which we will call hereafter second
step cascading:
the interaction of wounded nucleon from one (target or projectile) nucleus
with another nucleon from the same nucleus before the hadronization occurs.

The great influence of second step cascading process is straightforward.
On the
projectile side (high laboratory energy) the excitation of one initially
untouched nucleons by a excited state going backward in anti-laboratory
frame of reference leads to transfer a part of nucleon energy
to the secondary produced particles. Their energy in the nucleus center of
mass system are rather small but after transformation to the laboratory system
the effect is expected to be quite considerable.

Results presented in the present paper are obtained using the
geometrical chain model.
Recently \cite{gmc} the geometrical multichain (GMC) extension
to very high energies of the geometrical two-chain model \cite{g2c}
has been presented. Obtained proton--proton
multiparticle production characteristics show that up to at least
SPS energies the underlying physics can be treated as competitive to that
used by Dual Parton-like models or relativistic jet (LUND-like)
idea.


The GMC model of the nucleon--nucleon (hadron--hadron)
collision differs from the others (DPM- or LUND-type models)
in a details of the treatment of the creation of ``wounded nucleons''
(excited states, chains, strings).
In the GMC model a phenomenological description of the chain creation is
used with a very few parameters to be adjusted directly to the soft
hadronic interaction data, instead of using structure functions approach.
The geometrization of the interaction picture is in the parametrization of
the multiparticle production process as a function of the impact parameter
of colliding hadrons.


The wounded nucleon picture is adopted for GMC nucleus--nucleus interactions.
The detailed four-dimensional space-time history of each nucleon in both
colliding nuclei is traced.
Each wounded nucleon in moving for the constant time measured in its
c.m.s. equal in the present calculation to 1 fm/$c$.
Calculation shows that this is enough for the wounded nucleon (in the most
of cases) to get out from its own nucleus before the hadronization occurs,
if the interaction energy in the laboratory frame of about 1 TeV/nucleon
or more.
This means that for high energies sometimes wounded nucleons excited to the
high masses collide with ordinary nucleons from its own nucleus. The GMC
procedure described in \cite{g2c}, without any changes, is used in such cases.


The number of nucleons participating in the inelastic nuclei collision
is related closely to the respective cross-section ratios
and studied extensively in nucleus--nucleus interaction examinations.
The commonly used approximation based on the point--nucleon optical
approximation gives
\begin{equation}
n_{A}={{A\sigma_{pB}} \over {\sigma_{AB}}}\ \ \ ,\ \ \ \ \
n_{B}={{B\sigma_{pA}} \over {\sigma_{AB}}}\ \ \ ,\ \ \ \ \
\label{nnnu}
\end{equation}
\noindent
where $n_{A}$ in the average number of participants from $A$ nucleus.
The second step cascading process obviously has to
increase of the number of excited nucleons in both colliding nuclei
what is clearly seen.
The effect is stronger on the projectile (heavier) nucleus side.
In Fig.~1 mean numbers of nucleons participating
in Fe--N collision for different interaction energies are given.
Dashed lines represent the approximation used in evaluation of
Eq.(\ref{nnnu}). This approach does not take into account the
second step cascading mechanism,
thus these
lines depict only the ``primarily wounded'' nucleons in each colliding
nucleus (interacting inelastically during
the passage of one nucleus through the other -- the first step cascading).
After this the second step cascading takes place. The primarily wounded
nucleons inside each nucleus moves back and, when they traverse its own
nucleus, they can excite nucleons which survives the ``primary'' collision.


The increase of the number of wounded nucleons has to lead to increase of the
inelasticity.
The inelasticity is defined
as a fraction of interaction energy transferred to the
secondary particles created in the interaction. The remaining energy
is carried by the ``leading particle'' and transported downward to
next subsequent interaction in the hadronic cascade.
In the present paper we use the following definition:
\begin{eqnarray}
K_{\cal NN} =
{{\langle { \rm Energy\ carried\ by\ produced\ secondaries }\rangle} \over
{ \rm Initial\ energy\ of\ the\ projectile\ nucleus}} =
\ \ \ \ \ \ \ \ \ \ \ \ \ \ \ \ \ \ \ \ \ \ \ \
\nonumber \\
 { E_{\rm in}  -  E_{\rm nuclei}  -  \left( E_{\rm proton} + E_{\rm neutron}
 -  E_{\rm antiproton} -  E_{\rm antineutron}\right)  \over
{E_{\rm lab}}} ,
\label{knn}
\end{eqnarray}
\noindent
where $E_{\rm nuclei}$ is the energy of all nuclei remaining from colliding
nuclei (if there are any). All energies are given in laboratory
system of reference.
The GMC model
inelasticity for $p$--air is presented in Fig.~2 by a thick
solid curve. (For the comparison inelasticities of the other models
discussed in Ref.\cite{GS93} are also presented.)
As one can seen, it is almost constant over a wide range of
interaction energies.

In this paper we want to exhibit the role of the second step cascading
process in the nuclei interactions.
The quantitative evidence for the change in
inelasticities between the model with and without this
process is given in Fig.~3,
where the inelasticities for iron--air interactions are presented.



The depth of the position of the shower longitudinal development
curve $x_{\rm max}$ is one of the indisputable observable related
directly to the energy degradation rate in EAS.
The GMC model with the nucleus--nucleus interaction mechanism
described above was introduced to the
CORSIKA code \cite{corsika} developed by the KASCADE group.
The respective
calculations were performed for primary protons and
iron nuclei. For the second case, the model with and without second
step cascading mechanism was used.
Results together with Fly's Eye and Yakutsk data \cite{xmaxdata}
are given in Fig.~4.

The role of the second step cascading process in clearly seen.
It is increasing slowly with energy giving the positions
of the shower maxima higher in the atmosphere of about 10 g~cm$^{-2}$
at 10$^{16}$ eV and 30 g~cm$^{-2}$ at 10$^{19}$ eV.
Thus, if without this process the experimental data show the
``heavier than iron'' mass of primary cosmic rays, then with the second step
cascading taken into account data points lay between the proton
and iron mass predictions. The elongation rate changes from 76 g~cm$^{-2}$
to 70 g~cm$^{-2}$ when the second step cascading is switched on for
pure iron (for protons is equal to 63 g~cm$^{-2}$). All these
values are in good agreement with the measured rates as can be seen in
Fig.~4.


It is important to note that this conclusion
is obtained using the interaction model which does not
introduce any ``extraordinary'' effects which starts to dominate
the interaction picture at ultra high energies. Our model for
proton--proton inelastic collision is, at this point, a very
``conservative'' one (see, e.g., Fig.~2).
The agreement with data is mainly a consequence
of the second step cascading process which has to be present
in high energy nucleus--nucleus interactions.
This mechanism leads to the significant increase on the
inelasticity for nuclei collisions without changing the hadron--hadron
interaction characteristics.




\vspace{.3cm}

\centerline{
\hspace{.8cm}
\psfig{file=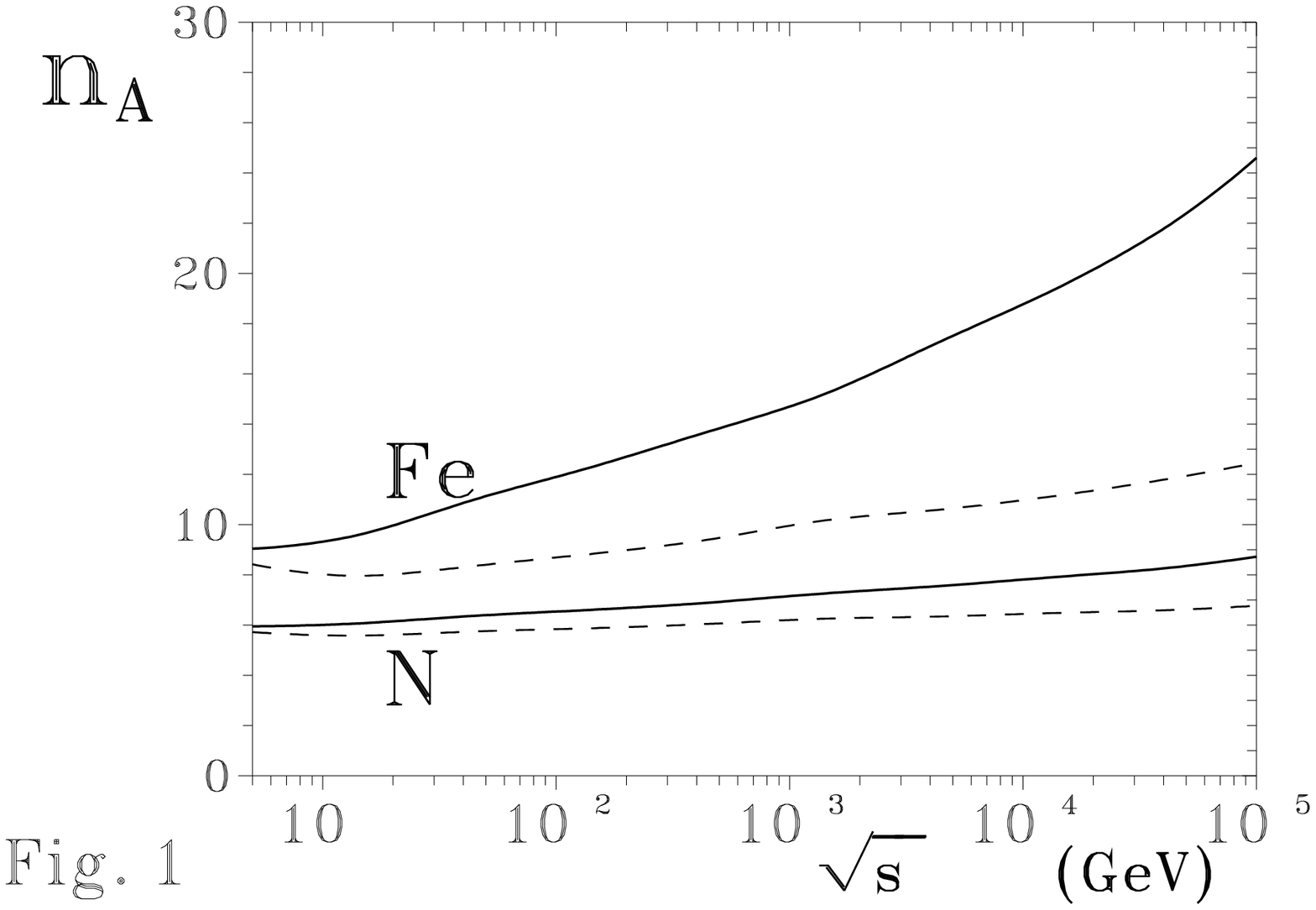,width=8.5cm}
\hspace{-1.cm}
%
\psfig{file=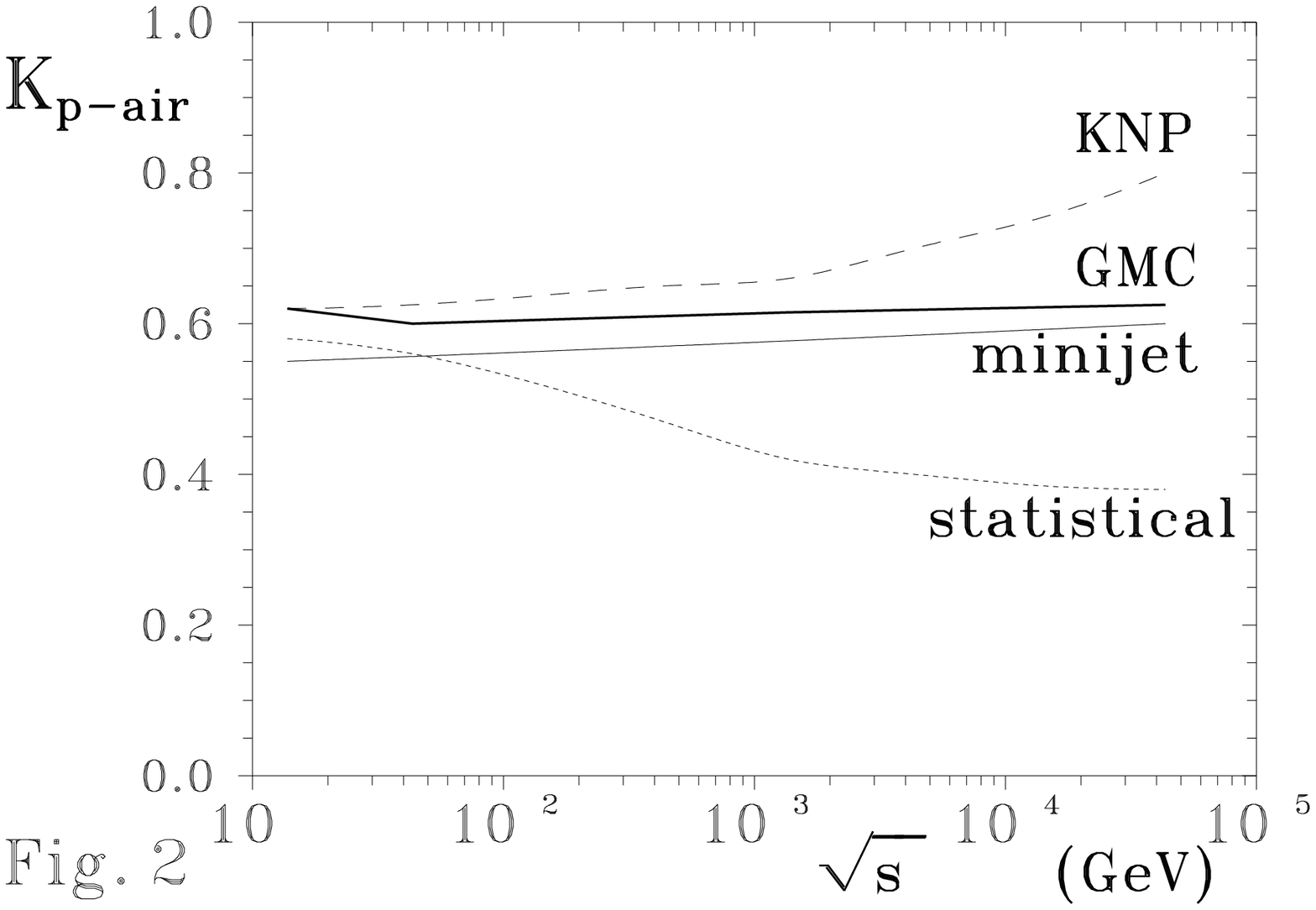,width=8.5cm}
}
\vspace{.5cm}
\noindent
\centerline{
\hspace{.8cm}
\psfig{file=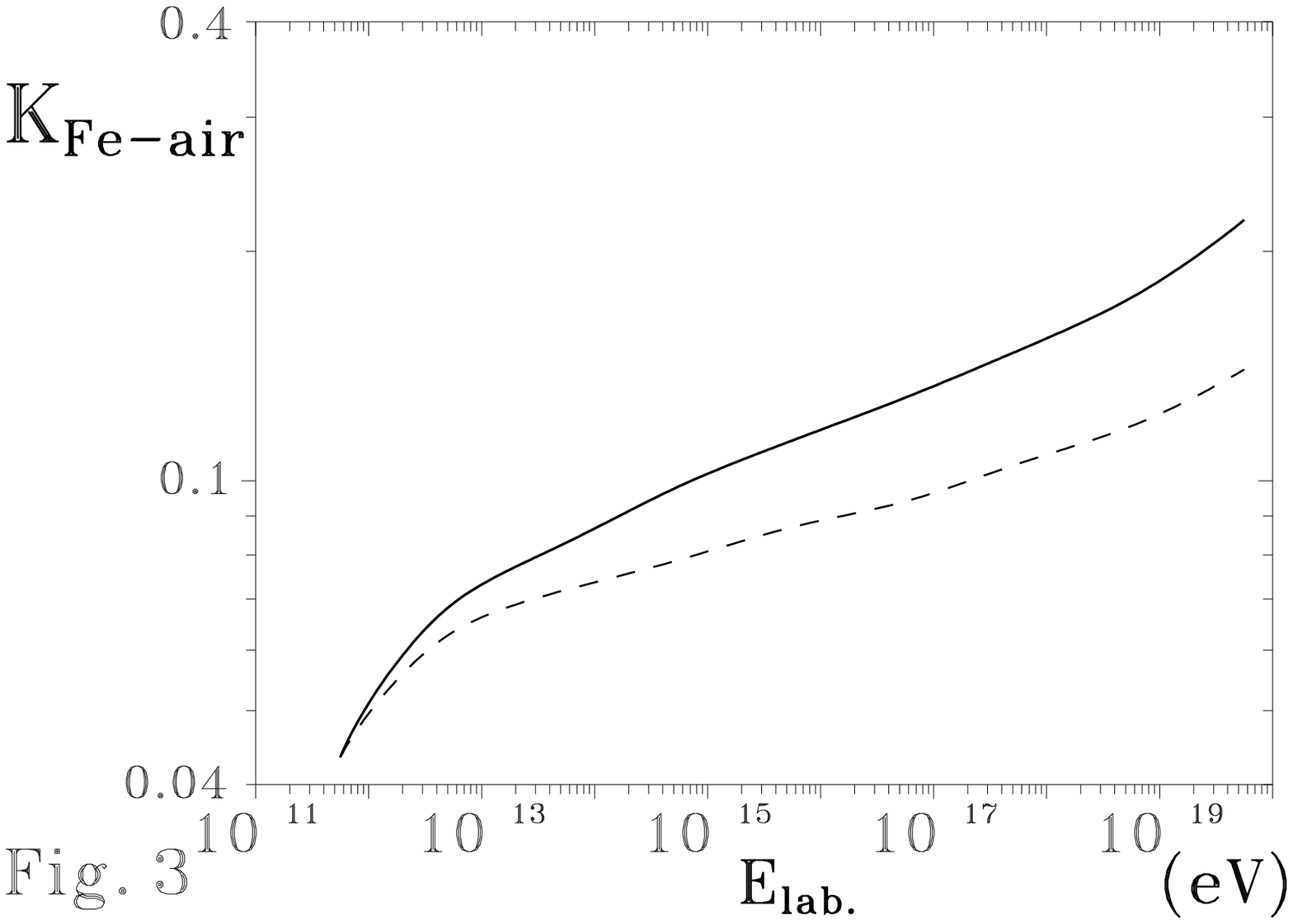,width=8.5cm}
\hspace{-1.cm}
%
%
\psfig{file=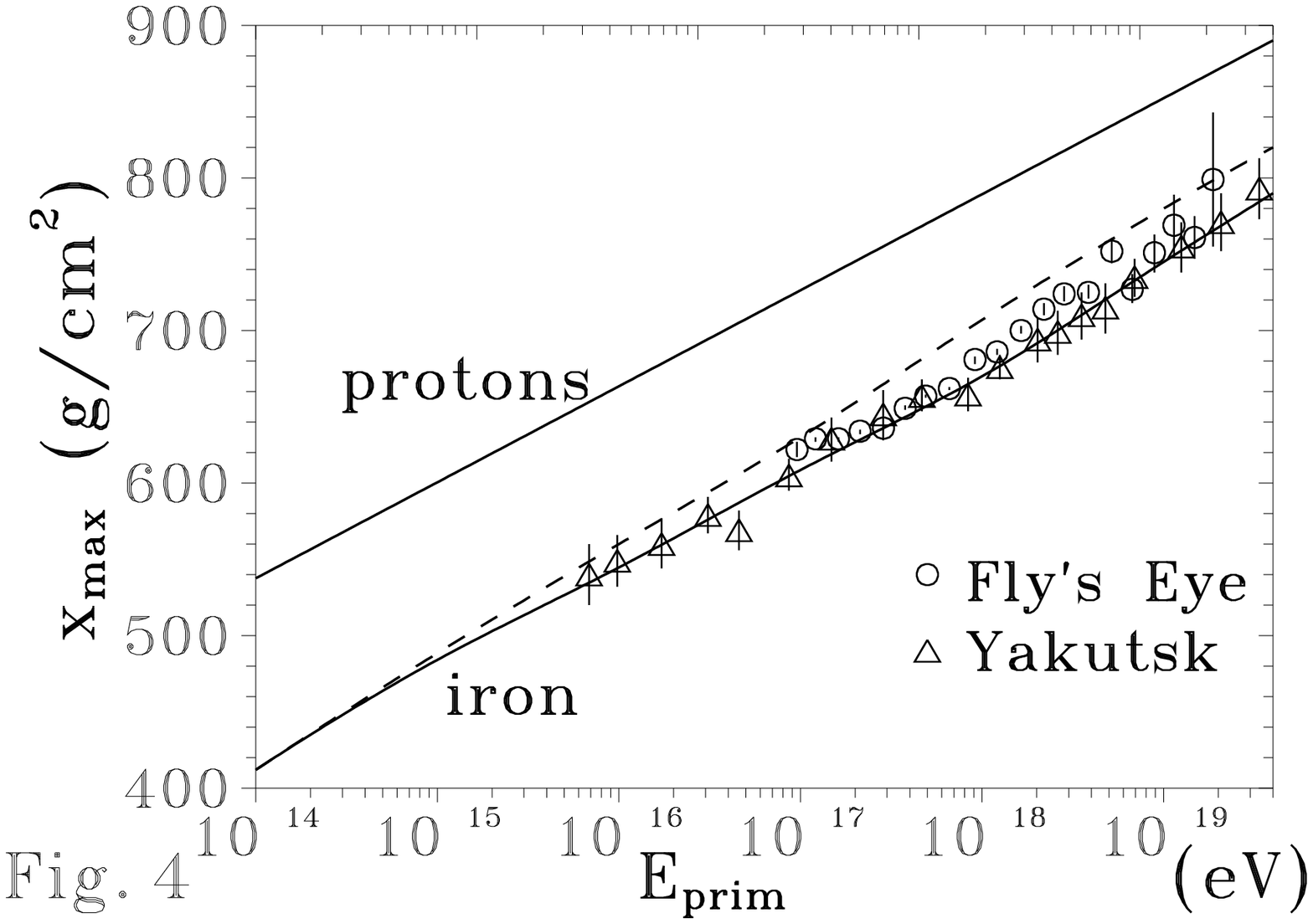,width=8.5cm}
}

\vspace{.8cm}

Fig.1.
{ Average wounded nucleon number for Fe--N interaction.
Two upper curves are for iron, lower -- for nitrogen nucleus respectively.
Dashed lines shows number of primary wounded nucleons
(before second step cascading), solid lines represent
final wounded nucleon numbers (after second step cascading process).}
\vspace{.3cm}

Fig.2.
{ Inelasticity coefficient of $p$--$air$ collisions.
Three models of $p$--$air$ used in Ref.~[4] analysis are presented
(thin lines) in comparison with the GMC model predictions (thick line).}
\vspace{.3cm}

Fig.3.
{ Inelasticity for $Fe$--$air$ collisions.
The solid and dashed lines in
were obtained with and without second step cascading
process taken into account.
for different laboratory energies.}
\vspace{.3cm}

Fig.4.
{ Depth of the sower maxima calculated with the GMC model compared with the
Fly's Eye and Yakutsk data. For the iron primaries the dashed line
represents the model without second step cascading process while
the solid one is obtained with this process taken into account.}

}

\end{document}